\pdfoutput=1

\documentclass[prl,aps,superscriptaddress,amsmath,amssymb,twocolumn]{revtex4-1}

\usepackage{braket,verbatim,bm,bbold,amsmath,amssymb,epsfig,float,color}
\usepackage[colorlinks,bookmarks=false,citecolor=blue,linkcolor=blue,hyperfootnotes=true,urlcolor=blue]{hyperref}

\newcommand{\bw}{\begin{widetext}}
\newcommand{\ew}{\end{widetext}}
\newcommand{\be}{\begin{equation}}
\newcommand{\ee}{\end{equation}}
\newcommand{\bea}{\begin{eqnarray}}
\newcommand{\eea}{\end{eqnarray}}

\definecolor{violet}{rgb}{0.62,0,1}
\definecolor{lightblue}{rgb}{0.12,0.56,1}
\definecolor{green}{rgb}{0.13,0.55,0.13}

\begin{document}

\title{Quasiparticle dynamics of symmetry resolved entanglement after a quench: \\the examples of conformal field theories and free fermions}  

\author{Gilles Parez}
\affiliation{Universit\'e catholique de Louvain, Institut de Recherche en Math\'ematique et Physique, Chemin du Cyclotron 2, 1348 Louvain-la-Neuve, Belgium
}

\author{Riccarda Bonsignori}
\affiliation{SISSA and INFN, Sezione di Trieste, via Bonomea 265, I-34136, Trieste, Italy}

\author{Pasquale Calabrese}
\affiliation{SISSA and INFN, Sezione di Trieste, via Bonomea 265, I-34136, Trieste, Italy}
\affiliation{International Centre for Theoretical Physics (ICTP), I-34151, Trieste, Italy}

\date{\today}

\begin{abstract}

The time evolution of the entanglement entropy is a key concept to understand the structure of a non-equilibrium quantum state. 
In a large class of models, such evolution can be understood in terms of a semiclassical picture of moving quasiparticles spreading the entanglement throughout the system.
However, it is not yet known how the entanglement splits between the sectors of an internal local symmetry of a quantum many-body system.
Here, guided by the examples of conformal field theories and free-fermion chains, we show that the quasiparticle picture can be adapted to this goal, leading to a 
general conjecture for the charged entropies whose Fourier transform gives the desired symmetry resolved entanglement $S_n(q)$. 
We point out two physically relevant effects that should be easily observed in atomic experiments: 
a delay time for the onset of $S_n(q)$ 
which grows linearly with $|\Delta q|$ (the difference from the charge $q$ and its mean value),
and an effective equipartition when $|\Delta q|$ is much smaller than the subsystem size. 

\end{abstract}

\maketitle

\paragraph{\it Introduction.---}\label{sec:intro} 

The time evolution of the entanglement in extended quantum systems starting from a non-equilibrium configuration became in the last decade a 
fundamental question with ramifications into many problems of contemporary physics, such as  
the equilibration and thermalisation of isolated many-body systems~\cite{PolkonikovRMP11, GE15, DKPR15, SI,EF16}, 
the emergence of thermodynamic entropy~\cite{C:18, DLS:13, SPR:11,ckc-14}, 
and the effectiveness of classical computers to simulate the quantum dynamics~\cite{SWVC:PRL, SWVC:NJP, PV:08, HCTDL:12, D:17}. 

These truly remarkable theoretical advances moved together with pioneering cold-atom and ion-trap experiments where it has been possible to directly 
measure the many-body entanglement of non-equilibrium quantum states \cite{kaufman-2016,exp-lukin,brydges-2018,brydges-2018,ekh-20}. 
In particular, in one of these experiments \cite{exp-lukin}, it has been recognized  that a more refined understanding of the many-body dynamics  
comes from the knowledge of how entanglement splits in different symmetry sectors.
Although the  {\it symmetry resolution of the entanglement} is the subject of an intense research activity of the last few years
\cite{GS,equi-sierra,bons,Luca,lr-14,cgs-18,fg-19,mdc-20b,fg-20,mdc-20,ccdm-20,tr-20,mrc-20,trac-20,Topology,Anyons,hc-20,as-20,bc-20b}, 
no  results are still available for the important case of a global quantum quench. 

In this Letter, we start filling this gap by initiating the study of the symmetry resolved entanglement after a global quantum quench in two paradigmatic
instances of many-body systems which are free fermions and conformal field theories (CFT). 
We will characterize these systems analytically and find new interesting effects that are expected to hold in more general circumstances, 
as it can be argued based on the quasiparticle picture for the entanglement spreading \cite{cc-05,ac-17,c-20}. 
Indeed, although we focus on rather simple models, we expect our findings to be qualitatively the same for large classes of systems. 
CFT and free fermion models have been the settings in which the study of the time evolution of many entanglement related quantities has 
been initiated (see e.g. \cite{cc-05,fc-08,ep-08,pp-07}) and only after many years generalized to more complex and realistic situations.


\paragraph{\it Quantities of interest. ---}
We consider an extended quantum system with an internal $U(1)$ symmetry, with conserved local charge $Q$.  
We take a bipartition $A\cup \bar A$ such that the charge $Q$ splits as $Q=Q_A+ Q_{\bar A}$. 
Consequently, the reduced density matrix $\rho_A$ satisfies $[\rho_A,Q_A]=0$, implying a block diagonal form, 
in which each block corresponds to an eigenvalue $q$ of $Q_A$, with normalized density matrix $\rho_A(q)$, i.e.
\begin{equation}
\label{rhoAblock}
\rho_A=\oplus_q \Pi_q \rho_A\Pi_q= \oplus_q [p(q)\rho_A(q)],
\end{equation}
where $\Pi_q$ is the projector on the subspace with eigenvalue $q$ and $p(q)= \mbox{Tr}(\Pi_q \rho_A)$ is the probability 
of having $q$ as outcome of a measurement of $Q_A$. 
The \textit{symmetry resolved R\'enyi entropies} are the entropies of the given sector, i.e.
\begin{equation}
\label{Snqdef}
S_n(q)=\frac{1}{1-n}\log \mbox{Tr}[\rho_A(q)^n],
\end{equation}
and for $n=1$ reduce to the von Neumann entropy $S_1(q)=-\mbox{Tr}[\rho_A(q)\log \rho_A(q)]$.
The latter satisfies the remarkable sum rule for the total entropy $S_1$ \cite{nc-10,exp-lukin}:  
\begin{equation}
\label{decompositionSvN}
S_1=\sum_q p(q)S_1(q)-\sum_q p(q)\log (p(q))\equiv S^c+S^n.
\end{equation}
The two terms above have been dubbed configurational ($S^c$) \cite{exp-lukin,wv-03,bhd-18} and number entanglement entropy ($S^n$) \cite{exp-lukin,kusf,kufs,ctd-19}.  
The former measuring the average of the entanglement in the charge sectors
and the latter the entropy due to the fluctuations of the charge within the  subsystem $A$.
In the experiment \cite{exp-lukin}, the time evolution of these entropies has been considered and it has been shown that, in the many-body localized phase,  
the number entropy grows quickly and soon saturates, irrespective of the strength of the interaction.
Conversely, the configurational entropy moves up very slowly, logarithmically in time, but only after a time delay, whose precise value depends on the interaction strength
and gets larger as the interaction decreases. 
These findings nicely explain former theoretical results for the total entanglement entropy \cite{bpm-12,Vosk2014}.

The computation of the symmetry resolved entanglement entropies from Eq. \eqref{Snqdef} requires the knowledge of the spectrum of $\rho_A$ and its resolution in $Q_A$, 
which is not easy because of the nonlocal nature of the projector $\Pi_q$. 
A more feasible path \cite{GS,equi-sierra} is based on the computation of the \textit{charged moments}
\begin{equation}
Z_n({\alpha})\equiv\mbox{Tr}[\rho_A^ne^{i\alpha Q_A}],
\end{equation}
whose Fourier transform
\begin{equation}
\mathcal{Z}_n(q)= \int_{-\pi}^{\pi}\frac{d \alpha}{2 \pi} e^{-i q \alpha} {Z}_n(\alpha)\equiv \mbox{Tr}[\Pi_q\rho_A^n],
\label{ZnqFT}
\end{equation}
readily provides the symmetry resolved quantities \eqref{Snqdef} as
\begin{equation}
S_n(q)=\frac{1}{1-n}\log \left[\frac{\mathcal{Z}_n(q)}{\mathcal{Z}_1(q)^n} \right].
\label{SvsZ}
\end{equation}
The probability $p(q)$ is just $p(q)=\mathcal{Z}_1(q)$.

\begin{figure}
\includegraphics[width=0.425\textwidth]{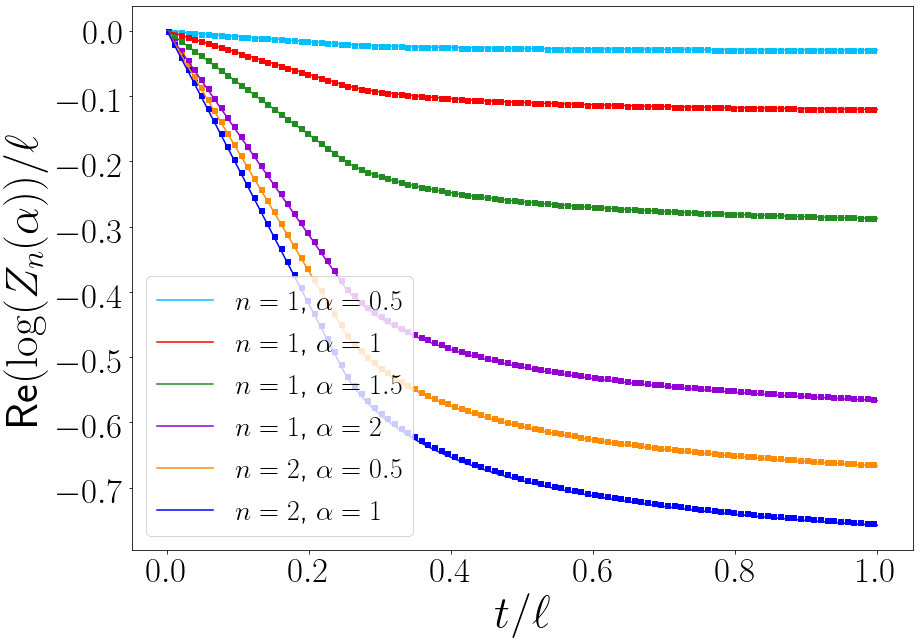}
\caption{The time evolution of the charged moments $Z_n(\alpha)$ after a quench from the N\'eel state in the free fermion model \eqref{HFF}.
We fix $\ell=120$ and plot as a function of $t/\ell$ for several values of $n$ and $\alpha$.
Numerical data (symbols) perfectly match the analytic prediction \eqref{Znamd} (full lines). 
} 
\label{FigZna}
\end{figure}

\paragraph{\it Free fermions. ---}
We study the evolution of the symmetry resolved R\'enyi entropies in the tight-binding model with Hamiltonian 
\begin{equation}
H=\sum_{i=1}^{L}(c_i^{\dagger}c_{i+1}+c_{i+1}^{\dagger}c_{i}),
\label{HFF}
\end{equation}
where ladder operators satisfy anticommutation relations $\{c_i,c_j^\dagger \} = \delta_{i,j}$ and $\{c_i,c_j \} = \{c_i^\dagger,c_j^{\dagger}\}=0$. 
The conserved charge is the fermion number $Q=\sum_{j} c^\dagger_j c_j$. 
The Jordan-Wigner transformation maps the model into the XX spin chain. 
We are interested in the entanglement of a block of $\ell$ consecutive sites in an infinite system. 
The reduced density matrix is obtained from the $\ell\times\ell$ matrix $C_A=\langle c_x^{\dagger}c_{x'} \rangle$ formed by the correlations 
with $x,x'\in A$ \cite{p-3,pe-09}.
Using standard algebra of Gaussian operators, the charged moments $Z_n(\alpha)$ can be written as \cite{GS}
\begin{equation}
\label{eq:Znalpha}
\log Z_n(\alpha)= \text{Tr} \log \left[ (C_A)^n e^{i \alpha}+ ( 1- C_A)^n \right].
\end{equation}

We begin our study with the quench from the N\'eel state $|N\rangle=\prod_{j=1}^{L/2}c_{2j}^{\dagger}|0\rangle$
(in spin language $|N\rangle\equiv |\uparrow\downarrow\uparrow\downarrow\uparrow\cdots\rangle$). 
We choose this initial state because it is the one engineered in most of the experiments \cite{exp-lukin,brydges-2018} and it is simple enough to allow full analytic
computations, serving as a guidance for the general case.
The correlation function is (see e.g. \cite{AC:19})
\begin{equation}
C(t)=\frac{\delta_{x,x'}}2 +\frac{(-1)^{x'}}2\int_{-\pi}^{\pi}\frac{dk}{2\pi}e^{ik(x-x')+4it\cos(k)}.
\end{equation}
For simplicity, we work with $\ell$ even.
The calculation of $Z_n(\alpha)$ in Eq. \eqref{eq:Znalpha} with the correlation matrix above can be performed 
in the space-time scaling limit, i.e. $t,\ell\to\infty$ with finite ratio, and it proceeds in full analogy to the case $\alpha=0$ presented in Ref. \cite{fc-08}. 
We expand the logarithm in Eq. \eqref{eq:Znalpha} in powers of $C_A$ to rewrite $\log Z_n(\alpha)$ as a series in ${\rm Tr}(C_A)^m$. 
These moments have been already calculated by multidimensional stationary phase technique \cite{fc-08}. The resulting power series can be summed up to obtain
\begin{equation}
Z_n(\alpha) = e^{i \ell \frac{\alpha}{2}} \left(\frac{\cos \frac{\alpha}{2}}{2^{n-1}} \right)^{\mathcal{J}} ,
\label{Znamd}
\end{equation}
where
\begin{equation}
\label{eq:J}
\mathcal{J} = \int \limits
\frac{d k}{2 \pi} \min[2v_k t, \ell], 
\end{equation}
with $v_k=2|\sin k|$. 
In Fig. \ref{FigZna} we compare this analytical prediction with ab-initio computations, finding perfect agreement. 

\begin{figure*}[t]
\begin{tabular}{l}
\includegraphics[width=0.425\textwidth]{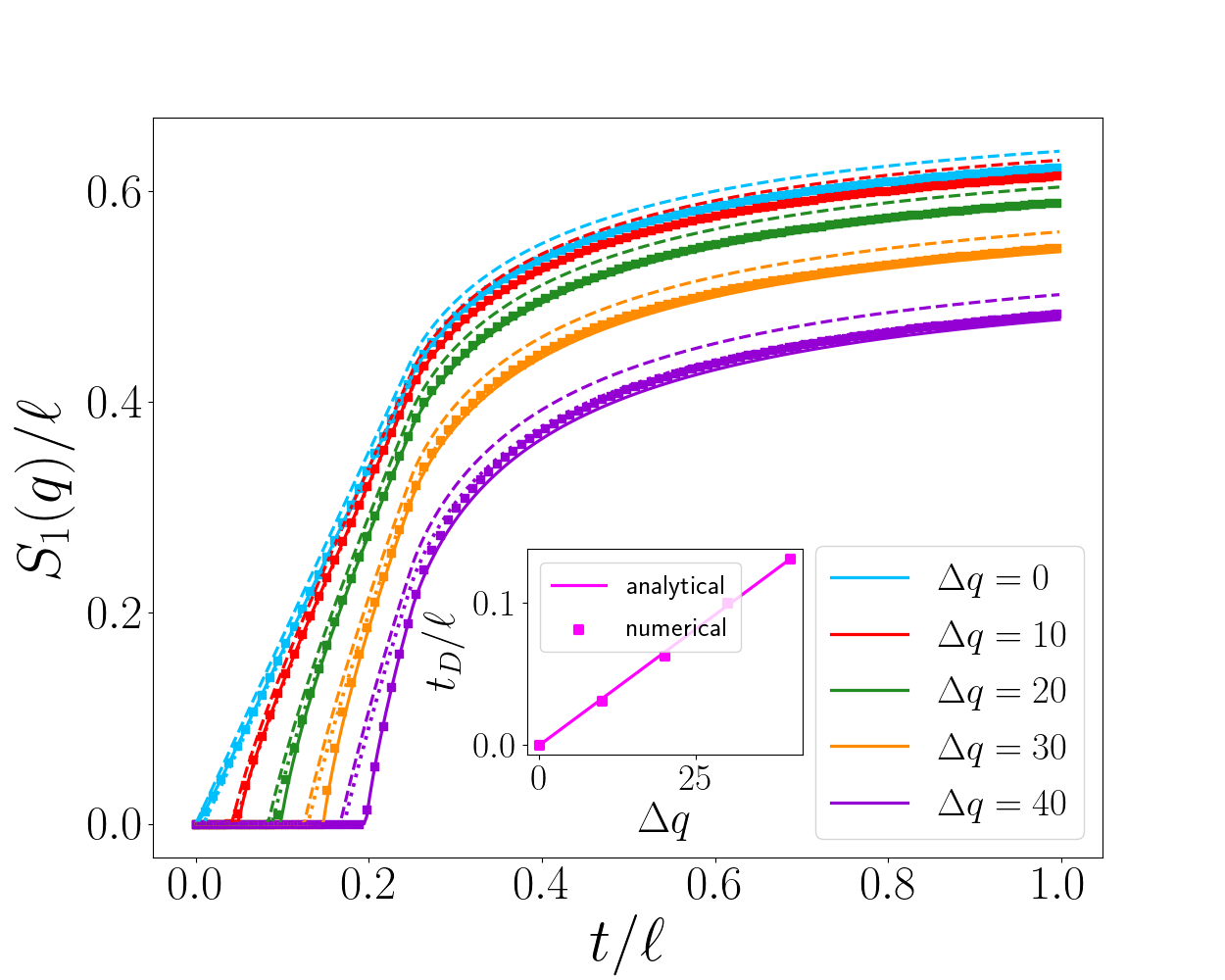} \qquad 
\includegraphics[width=0.425\textwidth]{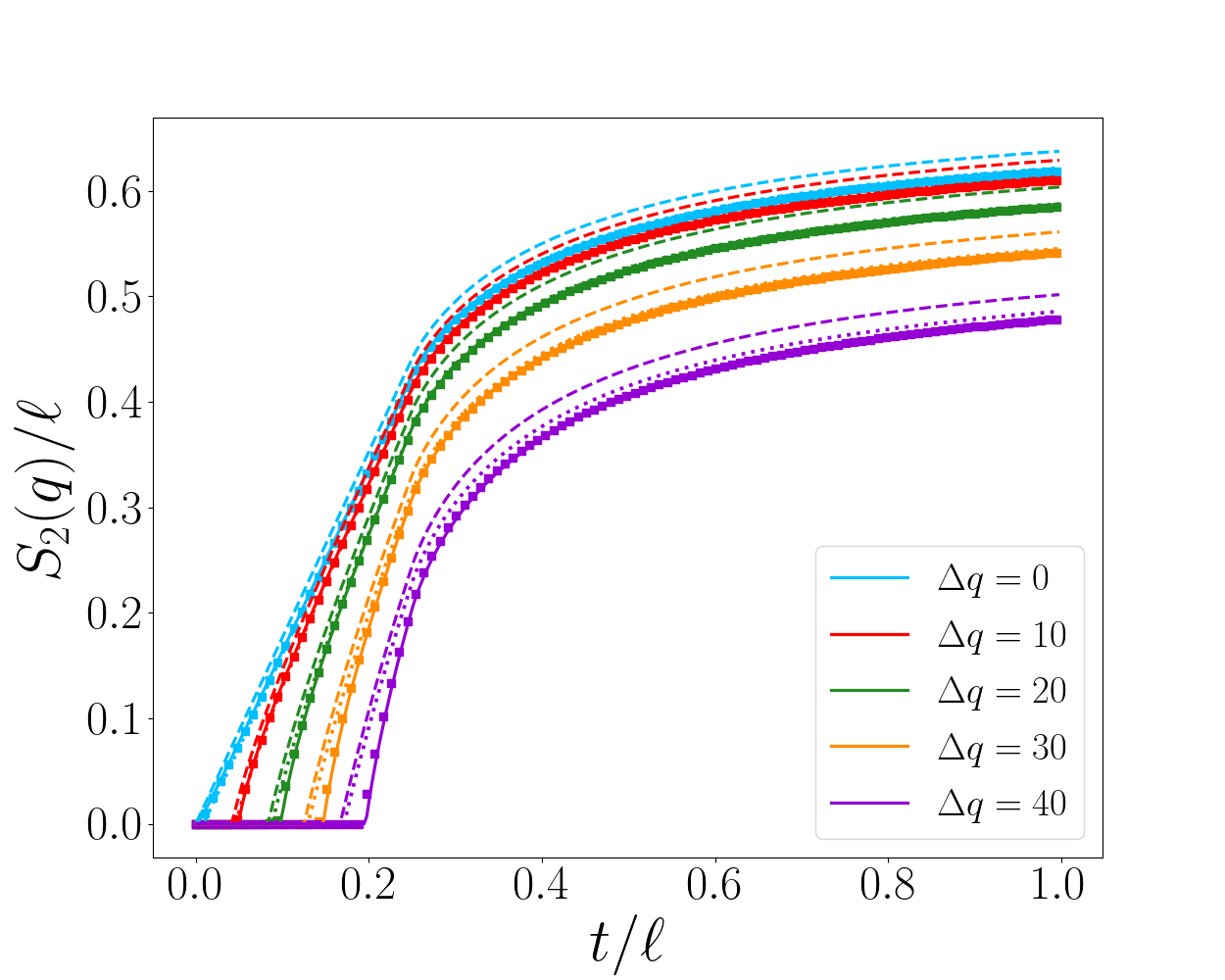} 
\end{tabular}
\caption{Time evolution of the symmetry resolved entanglement entropies $S_n(q)$ after a quench from the N\'eel state in the free fermion model \eqref{HFF}. 
The symbols are the exact numerical results for $\ell=160$ and various $\Delta q$. 
The full line is our prediction \eqref{eq:SnqIndep} with \eqref{eq:Z1qSPA} that 
perfectly matches the numerical data. Notice the delay $\propto \Delta q$ that is well captured by our prediction. 
The dashed line is the expansion for large $\ell$ and small $\Delta q$ while the dotted one is the same with logarithmic correction.
In the inset, we report the numerically calculated delays for $\ell=240$, obtained as the time when $S_1(q)/\ell=0.007$, and the analytic prediction $t_D=\pi |\Delta q|/4$. 
} 
\label{Fig:Snq}
\end{figure*}

The symmetry resolved moments for $q=\langle Q_A\rangle+\Delta q$, with $\langle Q_A\rangle=\ell/2$, are obtained plugging Eq. \eqref{Znamd} into \eqref{ZnqFT} to get 
\begin{equation}
\label{eq:Znq}
\mathcal{Z}_n(q)=2^{(1-n)\mathcal{J}} \int_{-\pi}^{\pi}\frac{d \alpha}{2\pi}  \left(\cos \frac{\alpha}{2} \right)^{\mathcal{J}} e^{-i\alpha\Delta q}.
\end{equation}
This integral can be evaluated analytically \cite{GR}, but for our aim it is convenient to use the saddle point method, which holds for large ${\cal J}$. 
We also assume $|\Delta q| \propto \ell$.
Since the dependence on $n$ in  Eq. \eqref{eq:Znq} is trivial, we focus on ${\cal Z}_1(q)$:
\begin{equation}
\label{eq:Z1qInt}
\mathcal{Z}_1(q)  =  \int_{-\pi}^{\pi} \frac{d \alpha}{2 \pi} e^{ \ell h(\alpha)}, \quad 
h(\alpha) = -i \alpha \frac{\Delta q}{\ell} + \frac{\mathcal{J}}{\ell} \log \cos \frac{\alpha}{2} .
\end{equation}
The saddle point is 
\begin{equation}
\label{eq:alphaStar}
\alpha^* = - 2i \, {\rm arctanh} \left( \frac{2  \Delta q}{\mathcal{J}}\right) \\
= i \log \left(\frac{\mathcal{J}-2\Delta q}{\mathcal{J}+2\Delta q}\right),
\end{equation}
that is purely imaginary only for ${\cal J}>2 |\Delta q|$, 
when we can deform the contour of integration to pass through $\alpha^*$ while staying in the region of analyticity of the integrand. 
Conversely for $\mathcal{J}<2 |\Delta q|$, $\alpha^*$ acquires a real part that leads to a non-zero  imaginary part of $h(\alpha^*)$ making ${\cal Z}_1(q)$ 
quickly oscillating in $\ell$ around $0$, a value to which it averages for all the relevant physics.
%
This is one of the main results of our Letter: {\it the symmetry resolved entanglement entropies start only after a delay time $t_D$ 
which grows linearly with $|\Delta q|$}. 
In fact, the equation $\mathcal{J}(t_D) = 2 |\Delta q|$ reads (as long as $2 v_M t_D< \ell$ self-consistently and $v_M\equiv \max v_k=2$) 
\begin{equation}
4 t_D \int_{-\pi}^{\pi} \frac{d k}{2 \pi}|\sin k| = 2| \Delta q| \quad  \Rightarrow \quad
t_D = \pi \frac{|\Delta q|}{4}. 
\label{tD}
\end{equation}
Instead, for $t>t_D$ (i.e. ${\cal J}> 2 |\Delta q|$), we have
\begin{equation}
\label{eq:Z1qSPA}
\mathcal{Z}_1(q) \approx e^{ \ell h(\alpha^*)} \sqrt{\frac{1}{2 \pi \ell |h''(\alpha^*)|}},
\end{equation}
that for large $\ell$ (i.e. large ${\cal J}$) is 
\begin{multline}
\log \mathcal{Z}_1(q)=-\left(\frac{\mathcal{J}}{2}+{|\Delta q| }\right)\log \left( 1+\frac{2 |\Delta q|}{\mathcal{J}}\right)\\ 
-\left(\frac{\mathcal{J}}{2 }-{|\Delta q|} \right)\log \left( 1-\frac{2 |\Delta q|}{\mathcal{J}}\right).
\end{multline}
From Eq. \eqref{SvsZ}, we can finally get the symmetry resolved entropies as 
\begin{equation}
\label{eq:SnqIndep}
S_n(q)= \mathcal{J} \log 2  + \log \mathcal{Z}_1(q).
\end{equation}
All these results are also found taking the large $\ell$ limit of the exact integral \eqref{Znamd} \cite{GR}, but the saddle point  
approach remains valid when $Z_n(\alpha)$ is not as simple as Eq. \eqref{Znamd}.
Interestingly, $S_n(q)$ does not depend on $n$.
Plugging Eq. \eqref{eq:Z1qSPA} into the above, one obtains the curves reported as full lines in Fig. \ref{Fig:Snq} that perfectly match the numerical data for $\ell=160$ and $n=1,2$. 
%
For $|\Delta q|\ll {\cal J}$ we have
\begin{equation}
\label{eq:SqCorr}
S_n(q)= \mathcal{J} \left( \log 2 - 2 \Big( \frac{|\Delta q|}{\mathcal{J}}\Big)^2\right), 
\end{equation}
reported as dashed lines in Fig.  \ref{Fig:Snq}. 
The deviations observed for small $|\Delta q|$ are due to the square root factor in Eq. \eqref{eq:Z1qSPA} that leads to a logarithmic correction to $S_n(q)$,
negligible in the limit of large $\ell$. This correction has been included in the dotted lines which match well the data as one moves away from $t_D$, to make 
${\cal J}$ significantly larger than $|\Delta q|$.
Eq. \eqref{eq:SqCorr} is another main result: {\it for small $|\Delta q|$ there is an effective equipartition of entanglement} \cite{equi-sierra} 
with violations of order  $(\Delta q)^2/\ell$.

To understand what happens for a more general initial state, we move to dimer state, i.e. a collection of neighbor singlets $|D\rangle=\bigotimes_{j} \left[\ket{\uparrow \downarrow}-\ket{\downarrow\uparrow}\right]_{2j-1,2j}$.
Following the same logic as before, with the correlation function, e.g., in Ref. \cite{f-14}, one obtains  after long, but simple, algebra
\begin{equation}
\label{eq:LogZnaDim}
\log Z_n(\alpha) = i\ell  \frac{\alpha}{2} + \int  \frac{d k}{2 \pi}{h}_{n,\alpha}(n_k) \min[2v_k t, \ell], 
\end{equation}
with $n_k=\frac{1+\cos k}2$, i.e., the mode occupation of the stationary state, and
\begin{equation}
{h}_{n,\alpha}(z)=  {\rm Re} \left[\log \left[ z^n e^{i \frac{\alpha}{2}}+ \left({1-z}\right)^n e^{-i\frac{\alpha}{2}}\right]\right].
\end{equation}
The main difference compared to the N\'eel quench is the function $h_{n,\alpha}(n_k)$ which is not constant. 
This modification makes the computation of ${\cal Z}_n(q)$ not feasible analytically. 
However, we can infer the validity of the two main physical results, namely the presence of a time delay $t_D\propto |\Delta q|$ and the effective equipartition for small $|\Delta q|$. 
Indeed, writing ${\cal Z}_n(q)\propto\int d \alpha e^{\ell g_n(\alpha)}$, the saddle point is given by $g'_n(\alpha^*)=0= 
-i \Delta q + i t\int \frac{dk}{2\pi}(\partial_{i\alpha} h_{n,\alpha}) 2v_k $.
Now, for $\alpha$ purely imaginary and arbitrary $n$, this last integral is a monotonic decreasing function of $i\alpha$ going from $4/\pi$ at $\alpha=-i\infty$
to $-4/\pi$ at $i\infty$. 
Hence the saddle point equation admits a purely imaginary solution only for $t> t_D= |\Delta q| \pi/4 $, which is the same delay time as for the N\'eel state.
It is straightforward to realize that the delay time is the same for any $n_k$ in Eq. \eqref{eq:LogZnaDim}  as long as $n_k\neq 0,1$ except in isolated points. 

Finally, proving equipartition for small $|\Delta q|$ is easy. It is enough to approximate $Z_n(\alpha)$ at Gaussian level as $Z_n(\alpha)\simeq Z_n(0) e^{-{\cal J}_n \alpha^2}$
to get immediately
\begin{equation}
\label{eq:SnqDimer}
S_n(q) = S_n- \frac{\Delta q^2}{4(1-n) } \left \{ \frac{1}{\mathcal{J}_n}-\frac{n}{\mathcal{J}_1}\right\},
\end{equation}
with $S_n$ the total entropy. It is slightly more complicated than Eq. \eqref{eq:SqCorr}, but physically equivalent.

\paragraph{\it Conformal field theory.---}
The CFT approach to global quantum quenches has been developed in Refs. \cite{cc-05,cc-06,cc-16}. 
The main idea is that the correlation functions at time $t$, following a quantum quench from a low entangled state $|\psi_0\rangle$, can be mapped to the path integral 
in a strip in Euclidean time of width $2\tau_0$ in which the operators are inserted at $\tau$ which must be analytically continued to $\tau\to \tau_0+i t$. 
The variable $\tau_0$ is an appropriate extrapolation length which, in some sense \cite{cc-06}, measures the distance of the initial state from an ideal 
conformally invariant boundary state, and hence the results are valid only for times $t$ and separations $\ell$ much larger than $\tau_0$.
The total R\'enyi entanglement entropies within this approach have been obtained \cite{cc-05} exploiting the fact that they are related to two-point correlators of properly defined 
twist fields. 
The very same ideas apply to the charged moments of a compact boson $\varphi$ (which includes free fermions), with conserved charge $Q_A=\int_A  dx \partial_x \varphi$. 
Indeed, the moments $Z_{n}(\alpha)$ are correlations of composite twist fields ${\cal T}_{n,\alpha}$ \cite{GS} of dimension 
$\Delta_{n,\alpha}= \frac{1}{12}(n-\frac1n) +\frac{K}{n} \frac{\alpha^2}{4\pi^2}$, with $K$ related to the compactification radius. 
Hence, the expectation value in the strip is the same as the one for the total moments $Z_n(\alpha=0)$ \cite{cc-05} with the change of the conformal dimension. 
Thus, after the analytic continuation, we have
\begin{equation}
Z_n(\alpha) = c_{n,\alpha} 
\left[  \frac{\pi^2}{4 \tau_0^2}  \frac{ \cosh \left(\frac{\pi \ell}{2 \tau_0}\right)+ \cosh\left( \frac{\pi v t}{\tau_0}\right)}{2 \sinh^2 \left( \frac{\pi \ell}{4\tau_0}\right) \cosh^2 \left(\frac{\pi vt}{2\tau_0}\right)}\right ]^{2\Delta_{n,\alpha}},
\end{equation}
with $c_{n,\alpha}$ a normalisation constant. 
For large $\ell/\tau_0$ and $t/\tau_0$, this simplifies to 
\begin{equation}
\label{eq:ZnaCFT}
\log Z_n(\alpha) =  \log Z_n(0) -  \frac{K \alpha^2}{4\pi n}  \frac{\min [2vt,\ell]}{ \tau_0}.
\end{equation}
This result resembles free fermions ones \eqref{Znamd} and \eqref{eq:LogZnaDim} with $\langle Q_A\rangle=0$ and with the difference that there is a single velocity $v$.
As a consequence of this single velocity and of the too simple dependence on $\alpha$ and $n$ in Eq. \eqref{eq:ZnaCFT}, 
the symmetry resolved moments and entropies do not show the time delay $t_D$ in Eq. \eqref{tD} and the functional form in $t,\ell$ is different. 
The conclusion is that while CFT captures {\it universal aspects of the charged entropy}, it expectedly fails to reproduce non-universal ones 
for the symmetry resolved quantities such as the time delay $t_D$.
Instead, the entanglement equipartition trivially follows from the Gaussian form of Eq. \eqref{eq:ZnaCFT}.

\paragraph{\it The general quasiparticle interpretation. ---} 
In integrable systems, the time evolution of the entanglement entropy can be understood in terms of the quasiparticle picture \cite{cc-05,ac-17}, 
in which quasiparticle excitations of opposite momentum are produced in pairs at time $t=0$ and then move ballistically through the system,
spreading entanglement and correlations.  
Assuming that the contributions of pairs of quasiparticles of momentum $\pm k$ to the charged entropy can be encoded in a single factor $f_{n,\alpha}(k)$, 
it naturally follows that
\begin{equation}
\log Z_n(\alpha) = i \langle Q_A\rangle \alpha +  \int  \frac{d k}{2 \pi}f_{n,\alpha}(k) \min[2v_k t, \ell], 
\label{QPf}
\end{equation}
where $v_k$ is the velocity of the entangling quasiparticles in the stationary state \cite{ac-17,bel-14}, $\langle Q_A\rangle$ 
the conserved mean charge within the subsystem, and the factor $\min[2v_k t, \ell]$ just comes from counting which pairs are shared between $A$ and $\bar A$ \cite{cc-05}. 
This form agrees and generalizes previous results  for free fermions  \eqref{eq:LogZnaDim} and CFTs \eqref{eq:ZnaCFT}. 
 Eq. \eqref{QPf} is written for a single species of quasiparticles, but the generalization to multiple ones is straightforward, since it just requires 
 to sum over all of them, as for the total entanglement \cite{ac-17}. 

Independently of the precise form of $f_{n,\alpha}(k)$,  we can infer the main general features of the symmetry resolved entanglement from Eq. \eqref{QPf}. 
The first one is the existence of the delay time $t_D$: 
in the saddle point calculation, $t_D$ is non-zero and proportional to $|\Delta q|$ if the domain of the derivative wrt $i\alpha$ of the integral 
in the rhs of Eq. \eqref{QPf} is finite for $\alpha$ purely imaginary. 
The precise value of $t_D$ depends on the details of the function $f_{n,\alpha}(k)$ and so on the specific quench.
In the quasiparticle picture, this delay can be physically understood as {\it the time needed to change the charge of an amount $|\Delta q|$ within the subsystem $A$}; e.g.,
having in mind a spin chain, this is the time to turn $|\Delta q|$ spins by local spin flips.
The other general result is the effective equipartition for small $|\Delta q|$ that follows from expanding Eq. \eqref{QPf} at quadratic order,
leading immediately to Eq. \eqref{eq:SnqDimer}.

Having understood the main qualitative features of the symmetry resolved entanglement, all the quantitative parts are encoded in the function $f_{n,\alpha}(k)$.
For free systems, we can either exactly solve the quench (as we did above) or we can infer it from the knowledge of the stationary state, 
conjecturing $f_{n,\alpha}(k)= h_{n,\alpha}(n_k)$ with $n_k$ the mode distribution in the stationary state (which is $n_k=1/2$ for the N\'eel initial state 
and $n_k=(1+\cos k)/2$ for the dimer state, matching all previous results).  
Anyhow, this last conjecture, although very reasonable, is not obvious at all, e.g., it is known \cite{gec-18} that the full counting statistics of the non-conserved order
parameter in the Ising model has contributions that are not captured by an equivalent of Eq. \eqref{QPf} and 
similar considerations have been drawn for the work statistics \cite{ppg-19}.
For genuinely interacting integrable models, 
for the total R\'enyi entropies (i.e. $Z_n(0)$) there are well known problems to reconstruct the time evolution \cite{ac-17a}, that can be circumvented close 
to $n=1$ \cite{ac-17}. The generalization to the charged entropy is in progress.
%


\paragraph{\it Conclusions. ---} In this Letter, we initiated the study of symmetry resolved entanglement after a quantum quench. 
Based on results from free fermions and CFTs, we conjectured the general formula \eqref{QPf} for the charged entropy which is expected to hold, within the 
quasiparticle picture, for arbitrary integrable models. From this general form two main physical results follow:
(i) The symmetry resolved entanglement with charge $|\Delta q|$ starts evolving only after a (calculable) time delay proportional to $|\Delta q|$;
(ii) For small $|\Delta q|$ there is an effective equipartition of entanglement broken at order $\Delta q^2/\ell$.

These findings however are the tip of an iceberg with ramifications into many branches of the quench dynamics. 
One important aspect is that our results allow us tackling (for free fermions quantitatively) 
more complicated entanglement measures such as mutual information and negativity \cite{ac-18b,AC:19}. 
Within the quasiparticle picture, the same remains true for the charged and symmetry resolved quantities.
Also, it is natural to wonder how to adapt the results to inhomogeneous initial states \cite{alba-inh,BFPC18,ABF19}.
Finally, it is a must to understand whether some of the techniques developed for non-integrable 
models \cite{NRVH:17,nvh-18,zn-20,BKP:entropy,GoLa19,bc-20,PBCP20,mac-20,cdc-18,fcdc-19,rpv-19,z-20}
apply to the symmetry resolved quantities and whether our main physical findings, time delay and equipartition, are robust.

\section*{Acknowledgments}
We acknowledge support from ERC under Consolidator grant number 771536 (NEMO).
GP thanks SISSA for hospitality and the Gustave B\"oel-Sofina Fellowships for financing his stay in Trieste. 
He is also supported by the Aspirant Fellowship FC 23367 from the F.R.S-FNRS and acknowledges support from the EOS contract O013018F.
PC is very grateful to Marcello Dalmonte for discussions on a closely related project and to Bruno Bertini, Andrea De Luca, and Lorenzo Piroli for 
comments and insights.



\begin{thebibliography}{999}


\bibitem{PolkonikovRMP11}
A.~Polkovnikov, K.~Sengupta, A.~Silva, and M.~Vengalattore,
\textit{Colloquium: Nonequilibrium dynamics of closed interacting quantum systems}, 
\href{http://dx.doi.org/10.1103/RevModPhys.83.863}{Rev. Mod. Phys. {\bf 83}, 863 (2011)}.

\bibitem{GE15}
C.~Gogolin and J.~Eisert, 
\textit{Equilibration, thermalisation, and the emergence of statistical mechanics in closed quantum systems}, 
\href{http://iopscience.iop.org/article/10.1088/0034-4885/79/5/056001}{Rep. Prog. Phys. {\bf 79}, 056001 (2016)}. 

\bibitem{DKPR15}
L.~D'Alessio, Y.~Kafri, A.~Polkovnikov, and M.~Rigol,
\textit{From Quantum Chaos and Eigenstate Thermalization to Statistical Mechanics and Thermodynamics}, 
\href{http://dx.doi.org/10.1080/00018732.2016.1198134}{Adv. Phys. {\bf 65}, 239 (2016)}.

\bibitem{SI}
P. Calabrese, F. H. L. Essler, and G. Mussardo, {\it Quantum Integrability in Out of Equilibrium Systems},
\href{http://iopscience.iop.org/article/10.1088/1742-5468/2016/06/064001}{J. Stat. Mech. (2016) 064001}. 

\bibitem{EF16}
F.~H.~L. Essler and M. Fagotti, 
\textit{Quench dynamics and relaxation in isolated integrable quantum spin chains}, 
\href{http://iopscience.iop.org/article/10.1088/1742-5468/2016/06/064002}{J. Stat. Mech. (2016) 064002}.











\bibitem{C:18}
P. Calabrese, \emph{Entanglement and thermodynamics in non-equilibrium isolated quantum systems}, 
\href{https://doi.org/10.1016/j.physa.2017.10.011}{Physica A {\bf 504}, 31 (2018)}.

\bibitem{DLS:13}
J. M. Deutsch, H. Li, and A. Sharma, \emph{Microscopic origin of thermodynamic entropy in isolated systems}, \href{https://doi.org/10.1103/PhysRevE.87.042135}{Phys. Rev. E {\bf 87}, 042135 (2013)}.

%

\bibitem{SPR:11}
L. F. Santos, A. Polkovnikov, and M. Rigol, \emph{Entropy of Isolated Quantum Systems after a Quench}, \href{https://doi.org/10.1103/PhysRevLett.107.040601}{Phys. Rev. Lett. {\bf 107}, 040601 (2011)}.

\bibitem{ckc-14}
M.~Collura, M.~Kormos, and P.~Calabrese, {\it Stationary entropies following an interaction quench in $1D$ Bose gas}, 
\href{https://doi.org/10.1088/1742-5468/2014/01/P01009}{J. Stat. Mech. P01009 (2014)}.



%
%
%
%
%
%
%
%
%
%
%
%
%
%
%



\bibitem{SWVC:PRL}
N. Schuch, M. M. Wolf, F. Verstraete, and J. I. Cirac, {\it Entropy Scaling and Simulability by Matrix Product States},
\href{https://doi.org/10.1103/PhysRevLett.100.030504}{Phys. Rev. Lett. {\bf 100}, 030504 (2008)}.

\bibitem{SWVC:NJP}
N. Schuch, M. M. Wolf, K. G. H. Vollbrecht, and J. I. Cirac, {\it On entropy growth and the hardness of simulating time evolution},
\href{https://doi.org/10.1088/1367-2630/10/3/033032}{New J. Phys. {\bf 10}, 033032 (2008)}.

\bibitem{PV:08}
A. Perales and G. Vidal, {\it Entanglement growth and simulation efficiency in one-dimensional quantum lattice systems},
\href{https://doi.org/10.1103/PhysRevA.78.042337}{Phys. Rev. A {\bf 78}, 042337 (2008)}.

\bibitem{HCTDL:12}
P. Hauke, F. M. Cucchietti, L. Tagliacozzo, I. Deutsch, and M. Lewenstein, {\it Can one trust quantum simulators?}
\href{https://doi.org/10.1088/0034-4885/75/8/082401}{Prog. Phys. {\bf 75} 082401 (2012)}.

\bibitem{D:17}
J. Dubail, \emph{Entanglement scaling of operators: a conformal field theory approach, with a glimpse of simulability of long-time dynamics in 1+1d}, \href{https://doi.org/10.1088/1751-8121/aa6f38}{J. Phys. A {\bf 50}, 234001 (2017)}.

\bibitem{kaufman-2016}
A.~M.~Kaufman, M.~E.~Tai, A.~Lukin, M.~Rispoli, R.~Schittko, P.~M.~Preiss, and  M.~Greiner, 
{\it Quantum thermalisation through entanglement in an isolated many-body system}, 
\href{http://dx.doi.org/10.1126/science.aaf6725}{Science {\bf 353}, 794  (2016)}.

\bibitem{exp-lukin}
A. Lukin, M. Rispoli, R. Schittko, M. E. Tai, A. M. Kaufman, S. Choi, V. Khemani, J. Leonard, and M. Greiner,
{\it Probing entanglement in a many-body localized system},
\href{https://science.sciencemag.org/content/364/6437/256/tab-figures-data}{Science {\bf 364}, 6437 (2019)}.

\bibitem{brydges-2018}
T.~Brydges, A. Elben, P. Jurcevic, B.~Vermersch, C. Maier, B. P. Lanyon,  P. Zoller, R. Blatt, and C. F. Roos,
  \textit{Probing entanglement entropy via randomized measurements},
 \href{http://dx.doi.org/10.1126/science.aau4963}{Science {\bf 364}, 260 (2019)}.
 
 \bibitem{ekh-20}
A. Elben, R. Kueng, H.-Y. Huang, R. van Bijnen, C. Kokail, M. Dalmonte, P. Calabrese, B. Kraus, J. Preskill, P. Zoller, and B. Vermersch,
{\it Mixed-state entanglement from local randomized measurements}, 
\href{https://arxiv.org/abs/2007.06305}{arXiv:2007.06305}.


\bibitem{lr-14}
N. Laflorencie and S. Rachel, {\it Spin-resolved entanglement spectroscopy of critical spin chains and Luttinger liquids},
\href{http://dx.doi.org/10.1088/1742-5468/2014/11/P11013}{J. Stat. Mech. P11013 (2014)}.

\bibitem{GS}
M. Goldstein and E. Sela, 
{\it Symmetry Resolved Entanglement in Many-Body Systems}, 
\href{https://journals.aps.org/prl/abstract/10.1103/PhysRevLett.120.200602}{Phys. Rev. Lett. {\bf 120}, 200602 (2018)}.

\bibitem{equi-sierra}
J. C. Xavier, F. C. Alcaraz, and G. Sierra, 
{\it Equipartition of the entanglement entropy}, 
\href{https://journals.aps.org/prb/abstract/10.1103/PhysRevB.98.041106}{Phys. Rev. B {\bf 98}, 041106 (2018)}.

\bibitem{cgs-18}
E. Cornfeld, M. Goldstein, and E. Sela,
{\it Imbalance Entanglement: Symmetry Decomposition of Negativity}, 
\href{http://dx.doi.org/10.1103/PhysRevA.98.032302}{Phys. Rev. A {\bf 98}, 032302 (2018)}.

\bibitem{bons}
R. Bonsignori, P. Ruggiero, and P. Calabrese, \textit{Symmetry resolved entanglement in free fermionic systems},
\href{https://iopscience.iop.org/article/10.1088/1751-8121/ab4b77}{J. Phys. A  {\bf 52}, 475302 (2019)}.

\bibitem{fg-20}
S. Fraenkel and M. Goldstein,
{\it Symmetry resolved entanglement: Exact results in 1d and beyond}, 
\href{http://dx.doi.org/10.1088/1742-5468/ab7753}{J. Stat. Mech. 033106 (2020)}.

\bibitem{fg-19}
N. Feldman and M. Goldstein, {\it Dynamics of Charge-Resolved Entanglement after a Local Quench}, 
\href{http://dx.doi.org/10.1103/PhysRevB.100.235146}{Phys. Rev. B {\bf 100}, 235146 (2019)}.


\bibitem{Luca}
L. Capizzi, P. Ruggiero, and P. Calabrese, \textit{Symmetry resolved entanglement entropy of excited states in a CFT },
\href{https://doi.org/10.1088/1742-5468/ab96b6}{J. Stat. Mech. (2020) 073101}.

\bibitem{mdc-20b}
S. Murciano, G. Di Giulio, and P. Calabrese, {\it Entanglement and symmetry resolution in two dimensional free quantum field theories},
\href{https://doi.org/10.1007/JHEP08(2020)073}{JHEP 2008 (2020) 073}.

\bibitem{mdc-20}
S. Murciano, G. Di Giulio, and P. Calabrese, {\it Symmetry resolved entanglement in gapped integrable systems: a corner transfer matrix approach}, 
\href{https://dx.doi.org/10.21468/SciPostPhys.8.3.046}{SciPost Phys. {\bf 8}, 046 (2020)}.

\bibitem{ccdm-20}
P. Calabrese, M. Collura, G. Di Giulio, and S. Murciano, {\it Full counting statistics in the gapped XXZ spin chain},
\href{https://doi.org/10.1209/0295-5075/129/60007}{EPL {\bf 129}, 60007 (2020)}.

\bibitem{tr-20}
M. T. Tan and S. Ryu,
{\it Particle number fluctuations, R\'enyi and symmetry-resolved entanglement entropy in two-dimensional Fermi gas from multi-dimensional bosonisation},
\href{https://dx.doi.org/10.1103/PhysRevB.101.235169}{Phys. Rev. B {\bf 101}, 235169 (2020)}.

\bibitem{mrc-20}
S. Murciano, P. Ruggiero, and P. Calabrese, {\it Symmetry resolved entanglement in two-dimensional systems via dimensional reduction},
\href{https://doi.org/10.1088/1742-5468/aba1e5}{J. Stat. Mech. (2020) 083102}.

\bibitem{trac-20}
X. Turkeshi, P. Ruggiero, V. Alba, and P. Calabrese,
 {\it Entanglement equipartition in critical random spin chains}, 
\href{https://doi.org/10.1103/PhysRevB.102.014455}{Phys. Rev. B {\bf 102}, 014455 (2020)}.

\bibitem{Topology}
K. Monkman and J. Sirker, {\it Operational Entanglement of Symmetry-Protected Topological Edge States},
\href{https://arxiv.org/abs/2005.13026}{arXiv:2005.13026}.

\bibitem{Anyons}
E. Cornfeld, L. A. Landau, K. Shtengel, and E. Sela, 
{\it Entanglement spectroscopy of non-Abelian anyons: Reading off quantum dimensions of individual anyons},
\href{http://dx.doi.org/10.1103/PhysRevB.99.115429}{Phys. Rev. B {\bf 99}, 115429 (2019)}.

\bibitem{hc-20}
D. X. Horv\'ath and P. Calabrese, {\it Symmetry resolved entanglement in integrable field theories via form factor bootstrap},
\href{https://arxiv.org/abs/2008.08553}{arXiv:2008.08553}.

\bibitem{as-20}
D. Azses and E. Sela, {\it Symmetry resolved entanglement in symmetry protected topological phases},
\href{https://arxiv.org/abs/2008.09332}{arXiv:2008.09332}.

\bibitem{bc-20b}
R. Bonsignori and P. Calabrese, {\it Boundary effects on symmetry resolved entanglement}, 
\href{https://arxiv.org/abs/2009.08508}{arXiv:2009.08508}.

\bibitem{cc-05}
P. Calabrese and J. Cardy, \emph{Evolution of Entanglement Entropy in One-Dimensional Systems}, 
\href{http://dx.doi.org/10.1088/1742-5468/2005/04/P04010}{J. Stat. Mech. (2005) P04010}.

\bibitem{ac-17} 
V.~Alba and P.~Calabrese, \emph{Entanglement and thermodynamics after a quantum quench in integrable systems}, 
\href{https://doi.org/10.1073/pnas.1703516114}{PNAS {\bf 114}, 7947 (2017)};\\
V. Alba and P. Calabrese, \emph{Entanglement dynamics after quantum quenches in generic integrable systems}, 
\href{https://doi.org/10.21468/SciPostPhys.4.3.017}{SciPost Phys. {\bf 4}, 017 (2018)}.

\bibitem{c-20}
P. Calabrese, {\it Entanglement spreading in non-equilibrium integrable systems}, 
Lectures for Les Houches Summer School on ``Integrability in Atomic and Condensed Matter Physics", 
\href{https://arxiv.org/abs/2008.11080}{arXiv:2008.11080}.

\bibitem{fc-08} M.~Fagotti and P.~Calabrese, \emph{Evolution of entanglement entropy following a quantum quench: Analytic results for the XY chain in a transverse magnetic field}, \href{https://doi.org/10.1103/PhysRevA.78.010306}{Phys. Rev. A {\bf 78}, 010306(R)}.

\bibitem{ep-08}
V. Eisler and I. Peschel, \emph{Entanglement in a periodic quench}, \href{https://doi.org/10.1002/andp.200810299}{Ann. Phys. (Berlin) {\bf 17}, 410 (2008)}.

\bibitem{pp-07}
T. Prosen and I. Pizorn, {\it Operator space entanglement entropy in a transverse Ising chain}, 
\href{https://doi.org/10.1103/PhysRevA.76.032316}{Phys. Rev. A {\bf 76}, 032316 (2007)}.

\bibitem{nc-10}
M.~A. Nielsen and I.~L. Chuang, \textit{{Quantum computation and quantum  information}}.
 \href{http://dx.doi.org/10.1017/CBO9780511976667}{Cambridge University Press, Cambridge, UK, 10th anniversary~ed. (2010)}.

\bibitem{wv-03}
H. M. Wiseman and J. A. Vaccaro, {\it Entanglement of Indistinguishable Particles Shared between Two Parties}, 
\href{https://doi.org/10.1103/PhysRevLett.91.097902}{Phys. Rev. Lett. {\bf 91}, 097902 (2003)}.

\bibitem{bhd-18}
H. Barghathi, C. M. Herdman, and A. Del Maestro, {\it R\'enyi Generalization of the Accessible Entanglement Entropy}, 
\href{https://doi.org/10.1103/PhysRevLett.121.150501}{Phys. Rev. Lett. {\bf 121}, 150501 (2018)};
H. Barghathi, E. Casiano-Diaz, and A. Del Maestro, {\it Operationally accessible entanglement of one dimensional spinless fermions},
\href{https://doi.org/10.1103/PhysRevA.100.022324}{Phys. Rev. A {\bf 100}, 022324 (2019)}.

\bibitem{kusf}
M. Kiefer-Emmanouilidis, R. Unanyan, J. Sirker, and M. Fleischhauer, {\it Bounds on the entanglement entropy by the number entropy in non-interacting fermionic systems},
\href{https://dx.doi.org/10.21468/SciPostPhys.8.6.083}{SciPost Phys. {\bf 8}, 083 (2020)};\\
M. Kiefer-Emmanouilidis, R. Unanyan, J. Sirker, and M. Fleischhauer, {\it Evidence for unbounded growth of the number entropy in many-body localized phases},
\href{https://dx.doi.org/10.1103/PhysRevLett.124.243601}{Phys. Rev. Lett. {\bf 124}, 243601 (2020)}.


\bibitem{kufs}
M. Kiefer-Emmanouilidis, R. Unanyan, M. Fleischhauer, and J. Sirker, {\it  Absence of true localization in many-body localized phases},
\href{https://arxiv.org/abs/2010.00565}{arXiv:2010.00565};
Y. Zhao, D. Feng, Y. Hu, S. Guo, and J. Sirker, {\it Entanglement dynamics in the three-dimensional Anderson model},
\href{https://arxiv.org/abs/2010.06678}{arXiv:2010.06678}.

\bibitem{ctd-19}
X. Cao, A. Tilloy, and A. De Luca, {\it Entanglement in a fermion chain under continuous monitoring},
\href{https://dx.doi.org/10.21468/SciPostPhys.7.2.024}{SciPost Phys. {\bf 7}, 024 (2019)}.




\bibitem{bpm-12}
J. H. Bardarson, F. Pollmann, and J. E. Moore,
{\it Unbounded growth of entanglement in models of many-body localization},
\href{http://dx.doi.org/10.1103/PhysRevLett.109.017202}{Phys. Rev. Lett. {\bf 109}, 017202 (2012)}.

\bibitem{Vosk2014}
R. Vosk and E. Altman, 
\emph{Dynamical Quantum Phase Transitions in Random Spin Chains},
\href{http://dx.doi.org/10.1103/PhysRevLett.112.217204}{Phys. Rev. Lett. {\bf 112}, 217204 (2014)}. 

\bibitem{p-3}
I. Peschel, {\it Calculation of reduced density matrices from correlation functions},
\href{http://dx.doi.org/10.1088/0305-4470/36/14/101}{J. Phys. A {\bf 36}, L205 (2003).}

\bibitem{pe-09}
I. Peschel and V. Eisler, {\it Reduced density matrices and entanglement entropy in free lattice models}, 
\href{https://iopscience.iop.org/article/10.1088/1751-8113/42/50/504003} { J. Phys. A {\bf 42}, 504003 (2009)}.


\bibitem{AC:19}
V. Alba and P. Calabrese, \emph{Quantum information scrambling after a quantum quench}, 
\href{https://doi.org/10.1103/PhysRevB.100.115150}{Phys. Rev. B {\bf 100}, 115150 (2019)}.


\bibitem{GR} 
Eq. \eqref{eq:Znq} is an integral representation of the reciprocal $\beta$ function, see \href{https://en.wikipedia.org/wiki/Beta_function}{Wikipedia}.

\bibitem{f-14}
M. Fagotti, {\it On conservation laws, relaxation and pre-relaxation after a quantum quench}, 
\href{https://doi.org/10.1088/1742-5468/2014/03/P03016}{J. Stat. Mech. (2014) P03016}.

\bibitem{cc-06}
P. Calabrese and J. Cardy, 
{\it Time-dependence of correlation functions following a quantum quench},
\href{https://doi.org/10.1103/PhysRevLett.96.136801}{Phys.\ Rev.\ Lett.\ {\bf 96} 136801 (2006)};\\
P. Calabrese and J. Cardy, 
{\it Quantum Quenches in Extended Systems},
\href{http://dx.doi.org/10.1088/1742-5468/2007/06/P06008}{J. Stat. Mech. P06008 (2007)}.

\bibitem{cc-16}
P. Calabrese and J. Cardy, 
\textit{Quantum quenches in 1+1 dimensional conformal field theories},
\href{http://iopscience.iop.org/article/10.1088/1742-5468/2016/06/064003}{J. Stat. Mech. (2016) 064003}.

\bibitem{bel-14}
L. Bonnes, F. H. L. Essler and A. M. La\"uchli, {\it  "Light-cone" dynamics after quantum quenches in spin chains},
\href{https://doi.org/10.1103/PhysRevLett.113.187203}{Phys. Rev. Lett. {\bf 113}, 187203 (2014)}.


\bibitem{ac-17a}
V. Alba and P. Calabrese, {\it Quench action and R\'enyi entropies in integrable systems},
\href{https://doi.org/10.1103/PhysRevB.96.115421}{Phys. Rev. B {\bf 96}, 115421 (2017)};\\
V. Alba and P. Calabrese, {\it R\'enyi entropies after releasing the N\'eel state in the XXZ spin-chain}, 
\href{http://dx.doi.org/10.1088/1742-5468/aa934c}{J. Stat. Mech. (2017) 113105};\\
M. Mestyan, V. Alba, and P. Calabrese, {\it R\'enyi entropies of generic thermodynamic macrostates in integrable systems},
\href{https://doi.org/10.1088/1742-5468/aad6b9}{J. Stat. Mech. (2018) 083104}.


\bibitem{gec-18}
S. Groha, F. H. L. Essler, and P. Calabrese, {\it Full Counting Statistics in the Transverse Field Ising Chain},
\href{http://dx.doi.org/10.21468/SciPostPhys.4.6.043}{SciPost Phys. {\bf 4}, 043 (2018)}.

\bibitem{ppg-19}
G. Perfetto, L. Piroli, and A. Gambassi, {\it Quench action and large deviations: Work statistics in the one-dimensional Bose gas},
\href{https://doi.org/10.1103/PhysRevE.100.032114}{Phys. Rev. E {\bf 100}, 032114 (2019)}.

\bibitem{ac-18b}
V. Alba and P. Calabrese, {\it Quantum information dynamics in multipartite integrable systems},  
\href{https://doi.org/10.1209/0295-5075/126/60001}{EPL {\bf 126}, 60001 (2019)}

\bibitem{alba-inh}
V.~Alba, \emph{Entanglement and quantum transport in integrable systems}, \href{http://dx.doi.org/10.1103/PhysRevB.97.245135}{Phys. Rev. B {\bf 97}, 245135 (2018)}. 

\bibitem{BFPC18}
B.~Bertini, M.~Fagotti, L.~Piroli, and P.~Calabrese, \emph{Entanglement evolution and generalised hydrodynamics: noninteracting systems}, \href{http://dx.doi.org/10.1088/1751-8121/aad82e}{J. Phys. A {\bf 51}, 39LT01 (2018)}.

\bibitem{ABF19}
V. Alba, B. Bertini, and M. Fagotti, \emph{Entanglement evolution and generalised hydrodynamics: interacting integrable systems}, \href{http://dx.doi.org/10.21468/SciPostPhys.7.1.005}{SciPost Phys. {\bf 7}, 005 (2019)}. 


\bibitem{NRVH:17}
A. Nahum, J. Ruhman, S. Vijay, and J. Haah, \emph{Quantum Entanglement Growth under Random Unitary Dynamics}, \href{https://doi.org/10.1103/PhysRevX.7.031016}{Phys. Rev. X {\bf 7}, 031016 (2017)}. 

\bibitem{nvh-18}
A. Nahum, S. Vijay, and J. Haah, \emph{Operator Spreading in Random Unitary Circuits}, \href{https://doi.org/10.1103/PhysRevX.8.021014}{Phys. Rev. X {\bf 8}, 021014 (2018)}.

\bibitem{zn-20}
T. Zhou and A. Nahum, \emph{The entanglement membrane in chaotic many-body systems}, 
\href{https://doi.org/10.1103/PhysRevX.10.031066}{Phys. Rev. X {\bf 10}, 031066 (2020)}.

\bibitem{cdc-18}
A. Chan, A. De Luca, and J. T. Chalker, {\it Solution of a minimal model for many-body quantum chaos},
\href{https://doi.org/10.1103/PhysRevX.8.041019}{Phys. Rev. X 8, 041019 (2018)}.

\bibitem{fcdc-19}
A. J. Friedman, A. Chan, A. De Luca, and J. T. Chalker, {\it Spectral statistics and many-body quantum chaos with conserved charge},
\href{https://doi.org/10.1103/PhysRevLett.123.210603}{Phys. Rev. Lett. {\bf 123}, 210603 (2019)}.


\bibitem{BKP:entropy}
B. Bertini, P. Kos, and T. Prosen, \emph{Entanglement Spreading in a Minimal Model of Maximal Many-Body Quantum Chaos},  \href{https://doi.org/10.1103/PhysRevX.9.021033}{Phys. Rev. X {\bf 9}, 021033 (2019)}.

\bibitem{GoLa19} 
S. Gopalakrishnan and A. Lamacraft, \emph{Unitary circuits of finite depth and infinite width from quantum channels},
\href{http://dx.doi.org/10.1103/PhysRevB.100.064309}{Phys. Rev. B {\bf 100}, 064309 (2019)}.

\bibitem{bc-20}
B. Bertini and P. Calabrese, {\it Prethermalisation and Thermalisation in the Entanglement Dynamics},
\href{https://doi.org/10.1103/PhysRevB.102.094303}{Phys. Rev. B {\bf 102}, 094303 (2020)}.

\bibitem{PBCP20} 
L. Piroli, B. Bertini, J. I. Cirac, and T. Prosen, \emph{Exact dynamics in dual-unitary quantum circuits},
\href{http://dx.doi.org/10.1103/PhysRevB.101.094304}{Phys. Rev. B {\bf 101}, 094304 (2020)}.

\bibitem{mac-20}
R. Modak, V. Alba, and P. Calabrese, {\it Entanglement revivals as a probe of scrambling in finite quantum systems}, 
\href{https://doi.org/10.1088/1742-5468/aba9d9}{J. Stat. Mech. (2020) 083110}.

\bibitem{rpv-19}
T. Rakovszky, F. Pollmann, and C. W. von Keyserlingk, {\it Sub-ballistic Growth of R\'enyi  Entropies due to Diffusion},
\href{https://doi.org/10.1103/PhysRevLett.122.250602}{Phys. Rev. Lett. 122, 250602 (2019)}.

\bibitem{z-20}
M. Znidaric, {\it Entanglement growth in diffusive systems},
\href{10.1038/s42005-020-0366-7.}{Comm. Phys. {\bf 3}, 100 (2020)};\\
T. Rakovszky, F. Pollmann, and C. W. von Keyserlingk, {\it Comment on ``Entanglement growth in diffusive systems''},
\href{https://arxiv.org/pdf/2010.07969.pdf}{arXiv:2010.07969}.

%


%
%
%
















%
%













%
%
%
%
%
%
%







\end{thebibliography}
\end{document}